\documentclass[manuscript,nonacm]{acmart}

\hypersetup{breaklinks=true}

\title{How to De-Escalate a Cyber Conflict}

\author{Robert Axelrod}
\orcid{https://orcid.org/0000-0003-4758-6590}
\affiliation{%
    \institution{University of Michigan}
    \department{School of Public Policy}
    \city{Ann Arbor}
    \state{MI}
    \postcode{48109}
    \country{USA}}
\email{axe@umich.edu}

\begin{abstract}
    De-escalation of a cyber conflict can be substantially more difficult than de-escalation of a conventional military conflict.
    This paper will first explain the reasons why de-escalation of a cyber conflict can be so difficult, and then present a list of suggestions about how to overcome these specific difficulties.
\end{abstract}

\begin{document}

\maketitle

\section{Introduction}

De-escalation of a cyber conflict can be substantially more difficult than de-escalation of a conventional military conflict.
This paper will first explain the reasons why de-escalation of a cyber conflict can be so difficult, and then present a list of suggestions about how to overcome these specific difficulties.
The goal is to help a cyber actor identify what needs to be done in advance of a conflict to attain the capabilities for de-escalation when desired, and what needs to be done during a crisis to actually achieve de-escalation.
In a future conflict between nations, cyber weapons may well be used in conjunction with kinetic, political and economic weapons.
For the purposes of this presentation, however, it is useful to focus solely on the cyber aspects of a conflict.

\section{The Challenges of Cyber De-Escalation}

\paragraph{Attribution}

The first problem for cyber de-escalation is that a victim of a cyber attack may not be sure who is launching the attack.

\paragraph{Uncertain Effects of Cyber Weapons}

Cyber weapons may have effects on the target far greater or far less than intended.
The effects may even be far beyond the intended target itself.
In contrast, when Japan attacked Pearl Harbor in 1941, Japan could not be very certain about how effective its aerial attack would be, but it could be fairly certain that the attack aimed at U.S.\ forces would not inadvertently get them into a conflict with the U.S.S.R.
The unpredictable effects of cyber weapons can not only make escalation hard to contain, but may also make de-escalation difficult to achieve.

\paragraph{Lack of Transparent Alert Status}

While surprise attacks are possible in conventional conflict, the target typically can monitor the alert status of a potential enemy.
For example, in the Cuban Missile Crisis of 1962, the Soviet Union was well aware that the U.S.\ had put its forces on high alert and was undertaking preparations for launching an invasion of Cuba.
So, when the alert level of these forces was lowered, it was a clear sign of de-escalation of tension.
However, most cyber capabilities do not require any visible forms of mobilization to reach full capacity, or have visible ways for outsiders to monitor a lowering of the side's cyber alert levels.
The lack of transparency on cyber alert status means that lowering their alert level may not be an effective method of communicating the intent to de-escalate.\footnote{An exception would be the cyber militias that involve large number of people whose alert status could not be hidden.}

\paragraph{Ambiguity of the Cyber Escalation Ladder}

In kinetic warfare, there is a widely shared sense of what counts as escalation and de-escalation.
For example, the use of a nuclear weapon would be seen by all parties as a very large jump in the escalation of combat.
In contrast, cyber conflict has no such clear breaks in its escalation ladder.
Even more importantly, two adversaries may have very different conceptions of what counts as a significant escalation, and therefore what cessation of activity would count as a significant de-escalation.
A meaningful example is that the U.S.\ might regard encouragement through cyber means of a succession movement as simply encouragement of free speech, whereas the target country might regard such encouragement as a major threat to the regime itself.

\paragraph{Weak Cyber Norms}

The most important cyber norms are based on the shared understanding that the laws of armed conflict apply to any sufficiently damaging cyber attack.
Beyond that, however, the lack of more specific norms governing cyber conflict make it difficult for a target to decide whether escalation or de-escalation would be an appropriate response to a given attack---especially if the attack were unprecedented in its methods, scope or effects.

\section{De-Escalation of a Cyber Conflict}

Guidance on how to de-escalate a cyber conflict comes in three forms: what your own side can do to de-escalate, how your side can make it easier for the other side to de-escalate, and what needs to be done in advance of a conflict to develop the capabilities needed to implement a de-escalation of a cyber conflict when desired.

\subsection{How to Implement Cyber De-Escalation in a Crisis}

\begin{enumerate}
    \item Reducing the level or scope of one's cyber activities is the most straightforward method of de-escalation of a cyber conflict.
    If the escalation itself was the result of tit-for-tat moves, then an effective way to signal the intent to de-escalate is to reply to the latest activity with a somewhat less intense response.\footnote{In the context of an iterated Prisoner's Dilemma, this is called ``generous tit-for-tat.''
    Indeed, in the presence of possible misunderstanding or misimplementation, generous tit-for-tat is an even more robust and effective strategy than plain tit-for-tat~\citep{Wu1995-copenoise}.}
    
    \item Taking care to communicate clearly by word or deed that one intends to de-escalate is especially important in a cyber conflict because of the five challenges to cyber de-escalation described earlier.
    
    \item Beware of spoilers.
    Because attribution is so difficult in a cyber conflict, there is always danger that an attempt at de-escalation can be undermined by the actions of third parties who want the conflict to continue.
    For example, third party spoilers in the form of non-state actors were able to undermine the Oslo Accords of 1973 between Israel and the Palestinian Liberation Organization.
    
    \item Beware of biased interpretation of ambiguous information.
    In the midst of any kind of conflict, there is a great risk that ambiguous information will be interpreted as evidence of further hostilities.
    For example, after the North Vietnamese navy attacked a U.S.\ destroyer in the Gulf of Tonkin on August 2, 1967, ambiguous indicators of a second attack two days later were taken by President Johnson to introduce the Gulf of Tonkin resolution that provided justification for U.S.\ escalation of the war in Vietnam.
    However, it is now clear the second attack probably never happened, but was merely the result of biased reading of ambiguous information~\citep{Hanyok1998-skunksbogies}.
    In the context of a cyber conflict, the danger of such biased reading of ambiguous information is especially great because the extent, nature, and even the source of a cyber attack can be hard to determine when time pressure is great.
\end{enumerate}

\subsection{How to Make It Easier for the Other Side to De-Escalate}

\begin{enumerate}
    \item Don't attack the other side's core values.
    For example, the North Korean cyber attack on Sony Entertainment was met with U.S.\ sanctions not only because of the economic damage the attack caused, but also because the attack explicitly aimed ``to threaten artists and other individuals with the goal of restricting their right to free expression''~\citep{Obama2015-imposingadditional}.
    
    \item Don't dehumanize the other side.
    
    \item Consider responding in a non-public manner in order to make it easier for the other side de-escalate.
    A report of the Defense Science Board suggests that a cyber attack could be designed to allow ``the possibility for quiet punishment known to the adversary leadership that does not `box them in' politically to a follow-on response''~\citep[p.~15]{Miller2017-taskforce}.
    
    \item Provide a costly signal of your intent to de-escalate~\citep{Fearon1994-domesticpolitical}.
    One kind of costly signal is something that makes it harder to mobilize your own public for continued hostilities.
    Examples of this kind of signal are showing respect for the other side by using the correct formal titles for their leaders and their nation, putting off unattainable demands for the indefinite future~\citep{Atran2008-reframingsacred}, and accepting responsibility for the part of the crisis that may be one's own fault.
    Precisely because these signals make it harder to sustain public support for continuing hostilities, they may be effective in signaling intent to de-escalate.
    
    \item Don't necessarily acknowledge that you have been the victim of a cyber attack.
    If the original cyber attack was not visible to the public, one way to achieve de-escalation is to not acknowledge the attack at all.
    For example, in 2007 Israel attacked a nuclear reactor under construction in Syria, Syria knew it had no effective response that wouldn't risk unacceptable escalation.
    So, Syria chose to not even acknowledge that it had been attacked.
    Likewise, there are circumstances in which the best response to a cyber attack is to avoid blaming the other side in the first place~\citep{Edwards2017-strategicaspects}.
\end{enumerate}

\subsection{How to Develop the Capabilities to Implement De-Escalation}

\begin{enumerate}
    \item The ability to de-escalate from a cyber conflict requires that cyber weapons be designed in a manner that allows them to be turned off.
    In other words, their autonomy should be limited not only in when they are turned on, but how they can be turned off.
    A corollary is that the control of cyber weapons should be centrally maintained in order to ensure that no one can interfere with the ability of national command authority to reduce or cease cyber maneuvers.
    
    \item The capacity to implement de-escalation of a cyber conflict depends not only on the proper design of one's cyber weapons, but also on the ability to stop allies, proxies and patriotic hackers from continuing or even escalating the conflict.
    
    \item Cyber weapons should be designed to avoid indignation on the part of the target that is so extreme that it would evoke a vengeful response.
    The problem with a vengeful response is that it may be taken with little regard to the cost or consequences, although it may feel ``rational'' and ``justified'' at the time.
    Needless to say, such an attack would make later de-escalation very difficult.
    Unfortunately, the very characteristics that tend to evoke a demand for moralistic vengeance may be present in a cyber attack: lack of warning, departure from precedent, ``cowardly'' in the sense that those who execute the attack are not in personal danger, and perfidious if disguised as coming from a non-combatant.
    Thus, the development of cyber capabilities and plans should take into account the risks of evoking moralistic vengeance that would make later de-escalation very difficult.
    
    \item Perhaps most important of all, communication channels between countries should be well constructed and continually available.
    For example, the U.S.\ and China agreed in 2008 to establish a hot line between the Pentagon and the Chinese defense ministry.
    But this mechanism has still not become fully operational.
    In August 2017, Gen.\ Joseph Dunford, the Chairman of the U.S.\ Joint Chiefs of Staff, met with his Chinese counterpart, Gen.\ Fang Fenghui.
    Afterward, Gen.\ Dunford said, ``We have ways of communicating. What we're looking for is a more responsive 24 hours a day, seven days a week communications link that can actually be used in a crisis''~\citep{Reuters2017-chinamilitary}.
    
    \item On a related point, a hot line should not be cut off to protest the other side's actions.
    Apparently, the Chinese side did this twice for extended periods of time to protest U.S.\ actions~\citep{Garnaut2013-wevalue}.
    While it is understandable that when one side feels disrespected, it may not wish to talk, but times of tension are exactly when a hot line may be most needed.
\end{enumerate}

\section{A Confidence Building Measure on De-Escalation}

Nations should make it part of their declaratory policy that their design of cyber weapons and their plans for potential use will include the capacity to de-escalate their effects once employed.

\bibliographystyle{ACM-Reference-Format}
\bibliography{ref}

\end{document}